\pdfoutput=1
\documentclass[aps,prl,twocolumn,groupedaddress,letterpaper]{revtex4}

\usepackage[pdftex]{graphicx}
\usepackage[usenames]{color}

\definecolor{darkblue}{rgb}{0,0.08,0.45}
\usepackage[bookmarksopen=true,bookmarksopenlevel=2,colorlinks=true,linkcolor=darkblue,anchorcolor=darkblue,citecolor=darkblue,urlcolor=darkblue,pdfstartview=FitH]{hyperref}

\begin{document}

\title{Micellar Crystals in Solution from Molecular Dynamics Simulations}
\author{J.A. Anderson$^{(a)}$, C.D. Lorenz$^{(b)}$ and A. Travesset$^{(a)}$}
\affiliation{(a) Department of Physics and Astronomy and Ames
Laboratory,\\ Iowa State University, Ames, Iowa 50011, USA. \\
(b) Materials Research Group, Engineering Division, King's
College, London Strand, London, WC2R 2LS, UK. }
\date{\today}

\begin{abstract}

Polymers with both soluble and insoluble blocks typically self-assemble into micelles, aggregates
of a finite number of polymers where the soluble blocks shield the insoluble ones from contact with
the solvent.  Upon increasing concentration, these micelles often form gels that exhibit crystalline
order in many systems. In this paper, we present a study of both the dynamics and the equilibrium
properties of micellar crystals of triblock polymers using molecular dynamics simulations. Our
results show that equilibration of single micelle degrees of freedom and crystal formation occurs
by polymer transfer between micelles, a process that is described by transition state theory. Near
the disorder (or melting) transition, bcc lattices are favored for all triblocks studied. Lattices with fcc ordering
are also found, but only at lower kinetic temperatures and for triblocks with short hydrophilic
blocks. Our results lead to a number of theoretical considerations and suggest a range of
implications to experimental systems with a particular emphasis on Pluronic polymers.

\end{abstract}

\maketitle

\section{Introduction}

Micelles of multi-block polymers are finite aggregates, typically around fifty polymers or less,
where the insoluble blocks shield the soluble ones from contact with the surrounding solvent.
Depending on control variables (temperature, polymer concentration, pH, etc..) micelles may
self-assemble into gels that exhibit long-range order such as bcc, fcc, hcp or other, more unusual,
crystals. Micellar crystals exhibit a number of unique properties that have made them extremely
attractive for fundamental studies as well as for applications \cite{Alexandridis2000,Likos2006}.
Extensively studied experimental systems include aqueous solutions of Pluronics (also known as
Poloxamers), ABA triblocks where A is Polyethylene oxide (PEO) and B is Polypropylene oxide
(PPO)~\cite{Wanka1994,Chu1995,Mortensen2001} and inverted Pluronics, where the A blocks are PPO and
the central block B is PEO~\cite{Mortensen1994,Mortensen2001} as well as non-aqueous systems such
as Polystyrene-Polyisoprene (PS-PI) diblocks in decane~\cite{McConnell1993} and other
solvents~\cite{Bang2002,Lodge2002,Lodge2004a,Lodge2004b}.

Micelles in solution are highly dynamical entities with polymers continually being absorbed and
released through time. Therefore, a micellar crystal has a considerably intricate structure, where
the long range order remains stable as the individual polymers are constantly hopping from one
micelle to the next. Theoretical approaches such as density functional or mean field theory
\cite{McConnell1996,vanVlimeren1999,Ziherl2001,Lam2003,Zhang2006,Likos2006,Grason2007} directly
study the ordered micelles and ignore the dynamical degrees of freedom of the polymers. Studies
using molecular dynamics (MD) have the advantage of providing a reasonably realistic description of
the dynamics, thus allowing the investigation of the role of single polymer degrees of freedom. In
contrast with other approaches, MD also offers the important advantage that no assumptions need to
be made about what is the possible thermodynamic state of the system.

The goal of this study is to predict phase diagrams of triblock polymers using MD simulations and
to gain an understanding of the dynamics of micellar crystal formation. Because of our ongoing
interest in Pluronic systems in aqueous solutions~\cite{Anderson2006}, we examine systems of
$A_nB_mA_n$ triblocks, where the $A$ beads are hydrophilic
and $B$ beads are hydrophobic. Although there have been a number of previous studies of multiblock
copolymers in solution using MD, see Ref.~\cite{Ortiz2006} for a recent review, the prediction of
crystalline structures presents substantial difficulties. Experimentally, it is well known that the
approach to thermodynamic equilibrium is slow in these systems, with time scales of the order of
minutes or hours. Therefore, even with suitably coarse-grained models, the long time scales
involved provide a considerable challenge for MD simulation studies.

In this paper, we provide a detailed investigation of self assembled micellar crystals using MD. We
aim to understand the mechanism by which they form, predict their range of stability and elucidate
their static and dynamic properties. We also present an in-depth study on the challenges associated
in reaching the thermodynamically stable state using MD and a successful strategy to overcome them.

\section{Model and Simulation Details}

\subsection{Simulation Details}

Systems of polymers are modeled by coarse-grained beads in an implicit solvent.
Ref.~\cite{Anderson2006} provides a more detailed justification than the outline provided here.
Individual polymers are $A_n B_m A_n$ symmetric triblocks, where $A$ beads are hydrophilic and $B$
beads are hydrophobic. All systems in this work are monodisperse with fixed values for $n$ and $m$.
Non-bonded pair potentials consist of a standard attractive Lennard-Jones potential for hydrophobic
interactions
\begin{equation}\label{Eq_U_bb}
    U_{\mathrm{BB}} = 4\varepsilon \left[ \left(\frac{\sigma}{r}
\right)^{12} - \left(\frac{\sigma}{r}\right)^6 \right] \ ,
\end{equation}
and a purely repulsive potential for hydrophilic interactions
\begin{equation}\label{Eq_U_ab}
    U_{\mathrm{AA,AB}} = 4\varepsilon \left(\frac{\sigma}{r}
\right)^{12} .
\end{equation}
Pair potentials are cutoff to zero at $r = 3.0\sigma$. All beads have the same mass $M$. $M$,
$\varepsilon$ and $\sigma$ are arbitrary and uniquely define the units of all numbers in the
simulations. Neighboring atoms in the polymer chain are connected together with a simple harmonic
potential. The packing fraction $\phi_P$ of the polymers is given by
\begin{equation}\label{Eq_density}
    \phi_P=\frac{\pi N_{\mathrm{poly}} N_{\mathrm{mon}}}{6(L/\sigma)^3} \ ,
\end{equation}
where $N_{\mathrm{poly}}$ is the number of polymers in a box of linear dimensions $L$ and $N_{\mathrm{mon}}=2n+m$ is the number of beads in each polymer. Simulation boxes are
cubic with periodic boundary conditions. For this work, we use a time step of $\Delta t = 0.005
\sqrt{m\sigma^2/\varepsilon}$. All simulations were performed using the LAMMPS software package
\cite{lammps} in the NVT ensemble via Nos{\'e}-Hoover dynamics \cite{Hoover1985}. 

The temperature sensitivity of Pluronic systems is a reflection of the underlying strong temperature dependence of the hydrophobic effect. To model temperature dependent phases in such systems, the solvent quality for the $A$ beads must change \cite{Anderson2006}. In this paper, we keep the solvent quality fixed and vary only the kinetic temperature when needed to equilibrate the systems properly.

Recently, a different implicit solvent model has been developed for describing Pluronic
systems~\cite{Bedrov2006}, where the pair potentials for the coarse-grained simulation are fitted
to results obtained from all-atom MD and quantum chemistry simulations. Although the
coarse-graining is different (each monomer $A$,$B$ represents one PEO or PPO monomer) than in our
model, the resulting potentials are quite similar to the ones in this work, Equations \ref{Eq_U_bb}
and \ref{Eq_U_ab}, with the only significant difference that the values of $\sigma$ are different
between the two $A$ and $B$ monomers, and $U_{\mathrm{AA}}$ shows a minor maximum.

\subsection{Observables}

A number of observable quantities are monitored for every recorded time step during a simulation
run. The micelles themselves are identified by the same algorithm used previously in
Ref.~\cite{Anderson2006}. Any hydrophobic beads within a distance of $r_{\mathrm{cut}} = 1.2
\sigma$ of one another are identified as belonging to the same micelle. Identified micelles
containing less than 3 polymers are typically free polymers in the process of being transferred
from one micelle to another and are removed from further consideration.  Observables such as
statistics of micelle aggregation number, gyration tensor, center of mass, and micelle lifetime
are calculated and examined for every simulation performed in this work. Methods used to calculate
these are described in Ref.~\cite{Anderson2006}.

The structure factor $S(\vec{q})$ is calculated over the center of mass coordinates $\vec{r}_i$ of
all $N_{\mathrm{mic}}$ micelles in the system using the formula
\begin{equation}\label{Eq_Sq}
    S(\vec{q})=C_0 \langle \left|  \sum_{i=1}^{N_{\mathrm{mic}}} e^{i\vec{q} \cdot \vec{r}_i} \right| ^2\rangle \ ,
\end{equation}
with the components of ${\vec{q}}$ as multiples of $\frac{2\pi}{L}$ due to the use of periodic boundary conditions.
The peaks in $S(\vec{q})$ are then used to reconstruct the full 3D real space
lattice basis, if it exists. In this manner, $S(\vec{q})$ is not being used to simulate real
scattering intensities as may be obtained by X-ray experiments, but as a mathematical order parameter
to discriminate between the different ordered structures that may be present. For convenience, the
normalization $C_0$ is chosen so that $S(\vec{q}=0)=1$.  A more sophisticated treatment that allows
continuum values of ${\vec q}$, suitable for quantitative comparisons with X-ray experiments, has
been recently introduced \cite{Schmidt2007}, but we do not use it here.

Individual polymers are constantly hopping from one micelle to another. This is quantified over the
entire simulation box as an overall rate of polymer transfer, $r_{\mathrm{PT}}$, by examining
contiguous simulation snapshots. Sets of indexed polymers belonging to each micelle are compared
between the snapshots to find the number of polymers transferred. The rate $r_{\mathrm{PT}}$ is
then expressed as a fraction of the number of polymers in the box transferred per one million
time steps. A polymer that is transferred out and back to the same micelle between snapshots will
not be counted by this analysis, so snapshots are recorded every 100,000 time steps to minimize
undercounting. At $k_B T/\varepsilon = 1.2$, a typical micelle only loses/gains one polymer per ten
snapshots recorded.

Radial distributions of the beads surrounding micelles are also of interest. These are calculated
by creating a histogram with bin width $dr$ and then counting the number of beads belonging to a
micelle $N_\mathrm{count}(r)$ that fall between $r$ and $r+dr$, where $r$ is the distance of the
bead from the center of mass of the micelle. Good statistics require averaging this histogram over
all micelles in the simulation and over all time steps after the micellar crystal has formed. The
average histogram is transformed into a radial density distribution of beads around a micelle by
calculating the packing fraction of beads in each bin
$$\phi_i(r) =  \frac{\pi }{6} \cdot \frac{\langle N_\mathrm{count}(r) \rangle}{4/3 \pi \left( \left(r+dr \right)^3 - r^3\right) / \sigma^3} \ ,$$
where $i$ refers to either $A$ or $B$ beads. We use $dr = 0.2\sigma$ to balance smooth graphs with the need for long simulation runs to obtain detailed statistics.

\section{Micellar crystals studied using MD}

There are two major challenges faced in using MD to determine equilibrium phases. First, the
simulation must last longer than any of the relaxation times in the system. Second, the simulation
box size must be chosen properly to avoid finite size effects. The first problem is related to the
kinetic temperature at which the system is run. The second problem can become particularly severe
for simulating crystals with three dimensional order, where the incorrect choice of an even large
box size $L$ can force the system into very distorted ordered phases. These two issues are
addressed systematically using simulations of the $A_{10}B_7A_{10}$ polymer. It provides
a coarse-grained description of one of the most extensively studied Pluronics,
F127~\cite{Wanka1994}. The conclusions of this study lead to a general methodology valid for
any other polymer.

\begin{figure}[hbt]
\includegraphics[width=8cm]{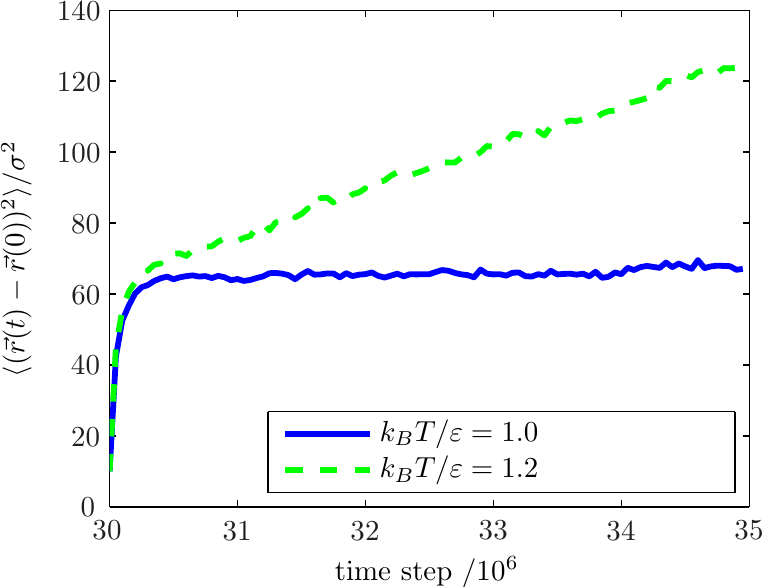}
\caption{\label{fig:polymer_msd} Mean squared displacement averaged over all  beads in the
simulation box. This data was obtained from the initial simulation runs of the $A_{10}B_7A_{10}$
polymer at $N_{\mathrm{poly}} = 500$ and $\phi_P = 0.20$. The origin of the time axis is 30 million
time steps which indicates that the recording of these results began after the system reached
equilibrium. In the case of $k_B T/\varepsilon = 1.0$ this equilibrium is a metastable state, and
the beads are not diffusing. The simulation run performed at $k_B T/\varepsilon = 1.2$ formed a fcc
lattice around time step 10 million which persisted until the end of the run at 35 million, and the
mean squared displacement shows a characteristic diffusive behavior.}
\end{figure}

\begin{figure}[hbt]
\includegraphics[width=8cm]{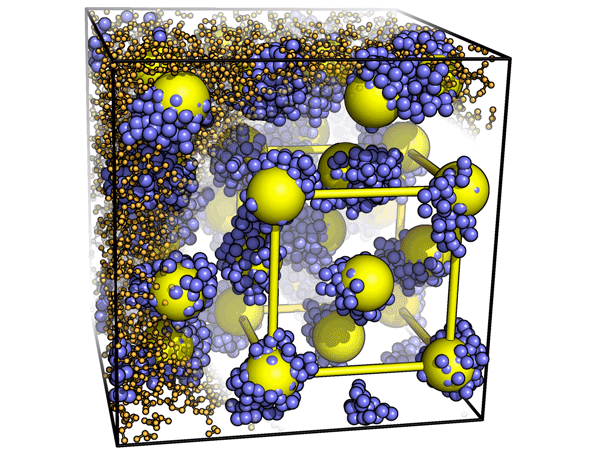}
\caption{\label{fig:fcc_500} Snapshot of a $A_{10}B_7A_{10}$  polymer simulation run at $\phi_P =
0.20$, $k_B T / \varepsilon = 1.2$, $N_{\mathrm{poly}} = 500$, taken after the fcc lattice formed.
$A$ beads are represented by orange spheres, and are shown with a reduced radius so they do not
obscure important details. Orange lines indicate bonds between these beads. $B$ beads are shown in
blue with a radius of $0.6\sigma$. Large yellow spheres are placed on the lattice reconstructed
from $S(\vec{q})$. Every yellow sphere is sitting on a micelle, visually confirming a perfect fcc
crystal. The $A$ beads have been removed around a single unit cell of the lattice and yellow lines
added to guide the eye. All snapshots are generated using PyMol~\cite{pymol}. }
\end{figure}

\begin{figure}[hbt]
\includegraphics[width=8cm]{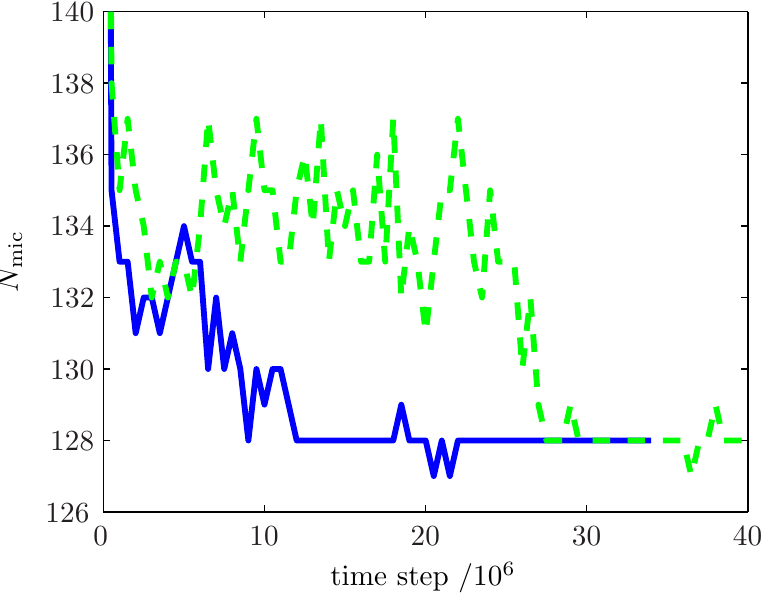}
\caption{\label{fig:n_mic_example} Examples of the behavior of $N_{\mathrm{mic}}$ during a simulation run of the $A_{10}B_7A_{10}$ polymer. Results are shown here for two independent runs with $\phi_P = 0.20$, $k_B T / \varepsilon = 1.2$ and $N_{\mathrm{poly}} = 2134$. The origin of the time axis (t=0) indicates the time step where the entire simulation started from a random initial configuration. Note how both systems eventually plateau at the same state $N_{\mathrm{mic}}=128$, though it occurs at different times. }
\end{figure}

\subsection{Micelle crystallization and kinetic temperature}

Initial simulations are performed with a system size of $N_{\mathrm{poly}} = 500$ at a kinetic temperature
$\frac{k_B T}{\varepsilon} = 1.0$ and with concentrations $\phi_P$ of $0.05$, $0.10$, $0.15$,
$0.20$ and $0.25$ run over $30$ million time steps each. In all cases, the initially randomly
placed $A_{10}B_7A_{10}$ polymers aggregate into micelles quickly in a few thousand time steps.

Visual examination indicates that concentrations at and above $\phi_P = 0.15$ are strong candidates
for micellar crystals. Micelles are locked in place and only move a few $\sigma$ from their average
positions. The analysis of the order parameter $S(\vec{q})$, however, indicates no long-range
ordered structures exist in any of these initial simulations. An inspection of the polymer transfer
shows that it remains negligible throughout all simulations. This is confirmed by calculating the mean squared displacement $\langle
(\vec{r}(t) - \vec{r}(0))^2 \rangle$ averaged over all beads in the system.
\autoref{fig:polymer_msd} clearly shows non-diffusive behavior at $k_B T/\varepsilon = 1.0$. These
results suggest that the individual micelles are quickly frozen in a configuration that is not
representative of equilibrium, thus preventing the entire system from reaching thermal equilibrium.

The lack of equilibration of micellar degrees of freedom suggests subsequent runs at a larger
kinetic temperature $k_B T/\varepsilon=1.2$. A single simulation run at $\phi_P = 0.20$ formed a
textbook fcc lattice after about 10 million steps, and is shown in \autoref{fig:fcc_500}.
Throughout the duration of the run, the rate of polymer transfer was substantial, about $7\%$ of
the polymers in the box every million time steps, \emph{even after} equilibrium is reached. Playing a movie of the simulation shows that after the lattice formed, micelles do not appear to move except by vibrating about their average positions. However, while the micelles appear static, polymers are
constantly being exchanged among the micelles, so that any individual polymer will eventually
explore the entire simulation box. This is independently confirmed by the analysis of the mean
squared displacement of beads in \autoref{fig:polymer_msd}, which shows a classic diffusion result
at $k_B T/\varepsilon=1.2$.

The behavior of the number of micelles $N_{\mathrm{mic}}$ in the box as a function of simulation
time is of particular interest. In simulation runs where micellar crystals are found, the system
reaches a plateau where $N_{\mathrm{mic}}$ remains constant, see \autoref{fig:n_mic_example} for an
example. Despite their dynamic character even at equilibrium, $N_{\mathrm{mic}}$ then remains
constant for the duration of the run. This correlation is a general feature in all simulation runs
performed. \emph{Every single} one that leads to a stable plateau in $N_{\mathrm{mic}}$ as
a function of time formed a micellar crystal confirmed by peaks in the order parameter $S(\vec{q})$.

\begin{figure}[hbt]
\includegraphics[width=8cm]{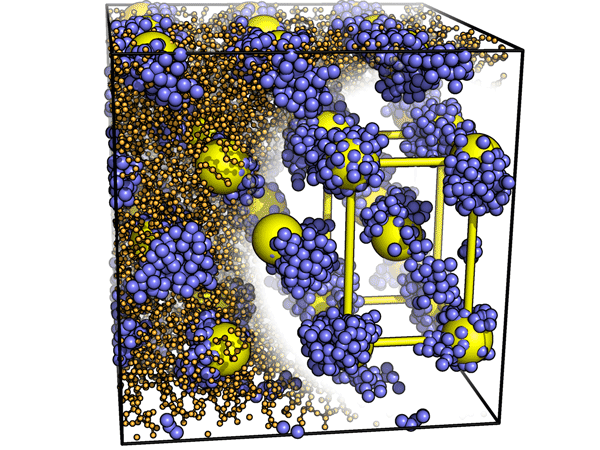}
\caption{\label{fig:distorted_600} Snapshot of a $A_{10}B_7A_{10}$ polymer simulation run at $\phi_P = 0.20$, $k_B T / \varepsilon = 1.2$, and
$N_{\mathrm{poly}} = 600$ taken after the lattice formed. The resulting lattice in this system is a body-centered tetragonal with $c/a = 1.5$. Coloring conventions are identical to \autoref{fig:fcc_500}}.
\end{figure}

\begin{table}[t]
\begin{tabular}{r|r|r|r|r}
\hline
No. configs & $N_{\mathrm{poly}}$ & $N_{\mathrm{mic}}$ & Ordering & $\langle N_{\mathrm{agg}} \rangle$ \\
1 & 500 & 32 & \emph{textbook} fcc & 15.625\\
3 & 500 & unstable &  none & \\
\hline
1 & 600 & 36 & none & 16.67 \\
3 & 600 & 36 & distorted bcc & 16.67 \\
\hline
5 & 800 & 48 & distorted bcc & 16.67 \\
\hline
5 & 900 & 54 & \emph{textbook} bcc & 16.67 \\
1 & 900 & 55 & distorted bcc & 16.36 \\
\hline
2 & 1000 & unstable & none & \\
4 & 1000 & 60 & distorted bcc & 16.67\\
\end{tabular}
\caption{\label{tab:N_test} Summary of results obtained from initial test runs.}
\end{table}

\begin{table}[ht]
\begin{tabular}{r|r|r|r|r}
\hline
No. configs & Targeted & $N_{\mathrm{poly}}$ & $N_{\mathrm{mic}}$ & Ordering \\
\hline
1 & 16 micelle bcc & 267 & 16 & \emph{textbook} bcc \\
\hline
4 & 108 micelle fcc & 1800 & unstable & none \\
\hline
4 & 128 micelle bcc & 2134 & 128 & \emph{textbook} bcc \\
\hline
3 & 250 micelle bcc & 4168 & unstable & none \\
\end{tabular}
\caption{\label{tab:algo_test_A10B7A10} Summary of simulation results from testing the algorithm
on $A_{10}B_7A_{10}$.}
\end{table}

\subsubsection{Box size effects}\label{sec:boxsize_effects}

The influence of the simulation box size choice is assessed by running additional simulations with $N_{\mathrm{poly}} = 500$, $600$, $800$, $900$, and $1000$ with fixed concentration $\phi_P = 0.20$ and $\frac{k_B T}{\varepsilon} = 1.2$. Several different random initial configurations are used at each system size to ensure repeatability. \autoref{tab:N_test} summarizes the results of all these simulation runs.

Interestingly, the additional three simulation runs at $N_{\mathrm{poly}} = 500$ do not exhibit fcc structures. Instead, each of them remains unstable with $N_{\mathrm{mic}}$ never achieving a constant plateau and $S(\vec{q})$ is devoid of peaks. Larger system sizes do reach stable ordered structures with constant $N_{\mathrm{mic}}$ and numerous peaks in $S(\vec{q})$. Most of the structures that occur appear bcc when examined visually, but a detailed analysis of the lattices indicates that many of them are distorted. One obviously distorted structure is depicted in \autoref{fig:distorted_600} where the lattice is body centered tetragonal with $c/a = 1.5$.

Other distortions include body centered tetragonal with $c/a = 1.054$ for $N_{\mathrm{poly}} = 800$ and a lattice that appears to be almost exactly bcc when $N_{\mathrm{poly}} = 1000$, except that the central micelle in the unit cell is shifted slightly from the true center. Lastly, a textbook bcc lattice is formed in the simulation runs with $N_{\mathrm{poly}} = 900$.

Examining the average aggregation number leads to a very illuminating result. It is found that for almost all simulations these numbers are \emph{identical}. This indicates that the average micelle aggregation number is independent of the box length (at a fixed $\phi_P$) and is only a function of the polymer structure, kinetic temperature and concentration.

Assuming without proof that either the fcc or the bcc lattice represents the real thermodynamic equilibrium of the system, the previous observation then suggests a way to generate \emph{magic numbers} of polymers to simulate equilibrium states free of finite size effects. In a bcc lattice, each cubic unit cell contains $C=2$ micelles, while the fcc lattice contains $C=4$. Therefore, in order to obtain a bcc or fcc lattice with $M$ by $M$ by $M$ unit cells in a cubic simulation box, the number of polymers needed to achieve this is given by \begin{equation}\label{Eq_MagicNumber} N_{\mathrm{poly}}=C M^3 \langle N_{\mathrm{agg}}\rangle . \end{equation}

\subsection{MD Simulations without finite size effects}\label{sec:algorithm}

An algorithm to simulate micellar crystals without finite size effects follows very naturally from the previous results, and is summarized in the following steps.

\begin{enumerate}

\item{\em Concentration selection:}  The concentration must be chosen high enough so that micelles pack into a potential crystalline state.

\item{\em Temperature selection: } The temperature should be chosen large enough to ensure a significant rate of polymer transfer, but low enough that the micellar crystals are not in a disordered phase. As a rule of thumb, we have been using a polymer transfer around $10\%$ of the polymers in the box every million steps.

\item{\em System size selection: } Calculate the average micelle number via test simulations and use \autoref{Eq_MagicNumber} to determine the final system sizes to perform simulations on.

\item{\em Ensure reproducibility: } The formation of micellar crystals is a stochastic process so several simulations with different initial configurations must be run. \end{enumerate}

The advantage of this algorithm is that steps 1-4 can be accomplished with relatively modest
computer resources on small system sizes, leaving the production runs with large polymer
numbers as the only computationally intensive calculations.

\subsection{Micellar crystals of general \texorpdfstring{$A_n B_mA_n$}{AnBmAn} triblocks }

\begin{figure}[bt]
\includegraphics[width=8cm]{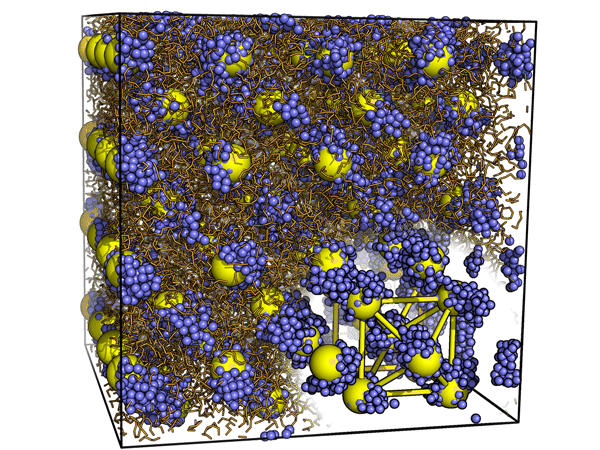}
\caption{\label{fig:bcc_2134} Snapshot of a $A_{10}B_{7}A_{10}$ simulation run at $\phi_P = 0.20$,
$k_B T / \varepsilon = 1.2$, and $N_{\mathrm{poly}} = 2134$, taken after the bcc lattice formed.
Coloring conventions are identical to \autoref{fig:fcc_500} }
\end{figure}

\subsubsection{ \texorpdfstring{$A_{10}B_7A_{10}$}{A10B7A10} }

Further simulation runs are carried out on the $A_{10}B_7A_{10}$ polymer  system to test the algorithm. These additional simulations target 16, 128 and 250 micelle bcc configurations, along with 32 and 108 micelle fcc ones. The simulations are summarized in \autoref{tab:algo_test_A10B7A10}. None of the simulation runs targeting fcc phases ever reach a stable number of micelles and correspondingly, $S(\vec{q})$ indicates there is no long-range order. On the other hand, both the 16 and 128 micelle bcc configurations are perfect and completely reproducible with the expected ordering confirmed by the structure factor. \autoref{fig:bcc_2134} shows a snapshot of the 128 micelle bcc phase after it forms. Further discussion on the calculated structure factors are presented below. Larger simulations attempting the formation of a 250 micelle bcc crystal never resulted in a stable $N_{\mathrm{mic}}$ plateau.

\subsubsection{ \texorpdfstring{$A_{20}B_{14}A_{20}$}{A20B14A20} }

The $A_{20}B_{14}A_{20}$ polymer is also studied as it provides a closer representation to the real
Pluronic F127, since there are twice as many monomers in a polymer and the level of coarse-graining
is less, as discussed in Ref.~\cite{Lam2003}. Applying the algorithm developed above to search for
micellar crystals, the concentration was selected at $\phi_P = 0.20$ and the kinetic temperature at
$k_B T / \varepsilon = 1.9$. Initial simulation runs establish that $\langle N_{\mathrm{agg}}
\rangle = 22.5 $. The results, summarized in \autoref{tab:algo_test_A20B14A20}, are similar to
those for the smaller $A_{10}B_{7}A_{10}$ polymer, with the bcc structure being the most commonly
found. The fcc phase again only appears in a single simulation run and is not reproducible.

\subsubsection{ \texorpdfstring{$A_6B_7A_6$}{A6B7A6} }

Given the prevalence of bcc lattices so far, a polymer with shorter  $A$-blocks ($A_6B_7A_6$),
expected to form more \emph{crew-cut} micelles and hence more prone to assemble into a fcc lattice,
is also investigated. Applying the algorithm developed above, the concentration is selected at
$\phi_P = 0.15$, the kinetic temperature at $k_B T / \varepsilon = 1.1$, and the average
aggregation number is found to be $\langle N_{\mathrm{agg}} \rangle = 13.875$. Again, the bcc
micellar crystal is most commonly found, as shown in the results in \autoref{tab:algo_test_A6B7A6}.
No fcc phases are found at this kinetic temperature.

However, the rate of polymer transfer at $k_B T / \varepsilon = 1.1$ is larger than $10\%$ per million steps, so some additional simulation runs are performed at a slightly reduced temperature $k_B T / \varepsilon = 1.07$, where polymer transfer is slightly reduced. In this case a completely reproducible fcc structure is found for relatively small system sizes, but no fcc lattices could be stabilized for larger ones as shown in the results in \autoref{tab:algo_test_A6B7A6_t07}.

\begin{table}[bt]
\begin{tabular}{r|r|r|r|r}
\hline
No. configs & Targeted & $N_{\mathrm{poly}}$ & $N_{\mathrm{mic}}$ & Ordering \\
\hline
2 & 16 micelle bcc & 360 & 16 & \emph{textbook} bcc \\
\hline
1 & 32 micelle fcc & 720 & 32 & \emph{textbook} fcc \\
1 & 32 micelle fcc & 720 & 35 & distorted bcc \\
2 & 32 micelle fcc & 720 & unstable & none \\
\hline
1 & 54 micelle bcc & 1215 & 54 & \emph{textbook} bcc \\
3 & 54 micelle bcc & 1215 & 55 & distorted bcc \\
\hline
2 & 108 micelle fcc & 2430 & unstable & none \\
\end{tabular}
\caption{\label{tab:algo_test_A20B14A20} Summary of simulation  results from testing the algorithm
on $A_{20}B_{14}A_{20}$ at $k_B T/\varepsilon = 1.9$.}
\end{table}

\begin{table}[bt]
\begin{tabular}{r|r|r|r|r}
\hline
No. configs & Targeted & $N_{\mathrm{poly}}$ & $N_{\mathrm{mic}}$ & Ordering \\
\hline
5 & 16 micelle bcc & 222 & 16 & \emph{textbook} bcc \\
\hline
5 & 32 micelle fcc & 444 & unstable & none \\
\hline
5 & 54 micelle bcc & 750 & 54 & \emph{textbook} bcc \\
\hline
5 & 108 micelle fcc & 1500 & unstable & none \\
\hline
3 & 128 micelle bcc & 1778 & 128 & \emph{textbook} bcc \\
1 & 128 micelle bcc & 1778 & 136 & distorted bcc \\
\hline
3 & 250 micelle bcc & 3472 & unstable & none \\
\hline
4 & 432 micelle bcc & 6035 & unstable & none \\
\end{tabular}
\caption{\label{tab:algo_test_A6B7A6} Summary of simulation results from testing the algorithm on
$A_{6}B_{7}A_{6}$ at $k_B T/\varepsilon = 1.1$.}
\end{table}

\begin{table}[bt]
\begin{tabular}{r|r|r|r|r}
\hline
No. configs & Targeted & $N_{\mathrm{poly}}$ & $N_{\mathrm{mic}}$ & Ordering \\
\hline
4 & 32 micelle fcc & 444 & 32 & \emph{textbook} fcc \\
1 & 32 micelle fcc & 444 & unstable & none \\
\hline
2 & 108 micelle fcc & 1500 & unstable & none \\
2 & 108 micelle fcc & 1500 & 103 & distorted bcc \\
1 & 108 micelle fcc & 1500 & 104 & distorted bcc \\
\end{tabular}
\caption{\label{tab:algo_test_A6B7A6_t07} Summary of simulation results from testing the algorithm
on $A_{6}B_{7}A_{6}$ after cooling to $k_B T / \varepsilon = 1.07$.}
\end{table}

\section{Dynamics of crystal formation}\label{sec:crystal_formation}

Molecular dynamics simulations not only allow the prediction of equilibrium phase diagram, but also
describe the dynamics of micelles as they evolve towards thermal equilibrium. In the simulations
presented previously, micelles quickly form from a completely random configuration of polymers, and
then after a long simulation time (10 to 20 million time steps) order into a micellar crystal. We
now present a quantitative picture of the dynamics of the polymers and micelles as the system
approaches equilibrium. All results in this section are obtained from the analysis of the
simulations of the $A_{10}B_{7}A_{10}$ polymer.

\subsection{Polymer transfer is an activated process}

\begin{figure}[bt]
\includegraphics[width=8cm]{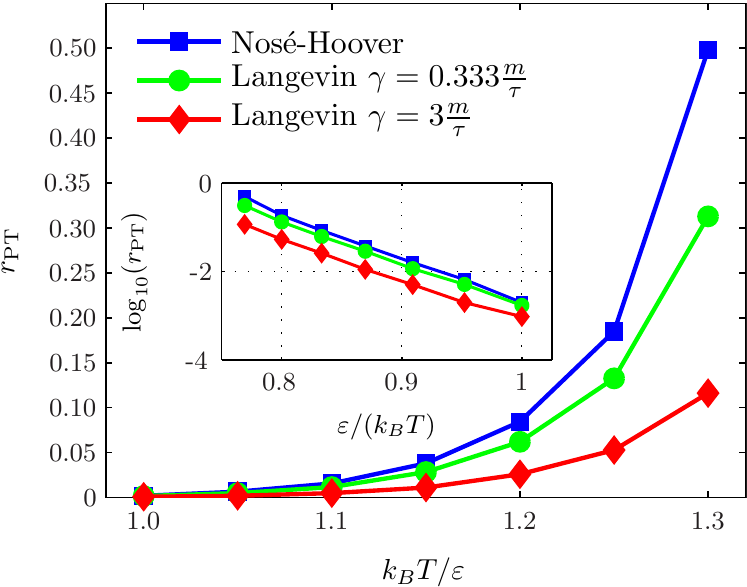}
\caption{\label{fig:poly_transfer} Polymer transfer $r_{\mathrm{PT}}$ versus temperature calculated
from simulation runs of the $A_{10}B_{7}A_{10}$ polymer at $\phi_P = 0.20$ and $N_{\mathrm{poly}} =
2134$. All simulations start from  an already equilibrated bcc phase. Results are included for the
Nos\'e-Hoover thermostat and Langevin thermostat with two different values of $\gamma$. The inset
plots the same data as a plot of $\log_{10}(r_{\mathrm{PT}})$ versus $\varepsilon / (k_B T)$ to
show that the slopes of the resulting lines ($-\Delta G^{\sharp}$) are universal.}
\end{figure}

Polymer transfer plays an important role in achieving the formation of micellar crystals in the
simulations discussed above. If there is too little, the single micelle degrees of freedom do not
reach equilibrium. Too much pushes the system into a disordered state. Moreover, the rate of
polymer transfer is extremely sensitive to the kinetic temperature. An equilibrated
$N_{\mathrm{poly}} = 2134$ bcc micellar crystal was taken as an initial configuration for
additional simulations that continued with $k_B T/\varepsilon$ ranging from $1.0$ to $1.3$.
\autoref{fig:poly_transfer} shows the results. The rate of polymer transfer $r_{\mathrm{PT}}$
starts near 0 at $k_B T/\varepsilon = 1.0$ and increases exponentially, following an Arrhenius
form
\begin{equation}\label{Eq:TST_micellar}
    r_{\mathrm{PT}}=r_O \exp(-\frac{\Delta G^{\sharp}}{k_B T}) \ .
\end{equation}
That is to say that polymer transfer is an activated process. Curve fitting, we find
\begin{equation}\label{Eq:G_dagger}
  \Delta G^{\sharp} \approx 10 \varepsilon \ .
\end{equation}

\autoref{fig:poly_transfer} also  includes results from simulations performed using a Langevin
thermostat~\cite{Frenkel2002}. It controls the temperature by adding an additional force to every
particle $\vec{F} = -\gamma \vec{v} + \vec{F}_{\mathrm{rand}}$ where the magnitude of the random
force $\vec{F}_{\mathrm{rand}}$ and $\gamma$ set the temperature through the fluctuation
dissipation theorem~\cite{Frenkel2002}. The results for two different values
of $\gamma$, which are plotted in \autoref{fig:poly_transfer}, show the dependence of the polymer
transfer for different drag coefficients.

In \autoref{fig:poly_transfer}, the slope of the lines in the inset plot are universal for both
thermostats and both values of $\gamma$. The universality of this value implies that the calculated
$\Delta G^{\sharp}$ is the free energy barrier between a polymer in a micelle to a transition state between micelles. While their slopes are universal, the y-intercepts (related to $r_O$) should depend on the diffusion coefficient of the hydrophilic beads, which, from the Einstein relation \cite{Doi1986} is inversely
proportional to the drag coefficient. This is reflected in the offsets of the various plots in
\autoref{fig:poly_transfer}, which are clearly different.

The difference in free energy between a polymer within a micelle and in the transition state entirely surrounded by solvent and hydrophilic beads can be roughly estimated as $\Delta G^{\sharp} \sim m_{exp} \cdot \varepsilon$, where $m_{exp}$ is the
number of hydrophobic beads exposed to solvent. If the length of hydrophobic block is low,
$m_{exp}\sim m$. Thus for the $A_{10}B_{7}A_{10}$ polymer analyzed here, it is expected that $\Delta G^{\sharp} \sim 7 \varepsilon$, which is consistent with the measured value. Hydrophobic beads in polymers with a longer hydrophobic block are expected to form a globule in the transition state, leading to a slower increase of $\Delta G^{\sharp}$ going as $m_{exp}\sim m^{2/3}$. Assuming that $r_O$ remains constant as $m$ increases, then maintaining the same rate of polymer transfer $r_{\mathrm{PT}}$ will require increasing the kinetic temperature by the same factor. These considerations agree remarkably with the kinetic temperature of $k_B T /\varepsilon = 1.9$ selected for the $A_{20}B_{14}A_{20}$ polymer simulations, 
$$\left( \frac{ m_{\mathrm{new}} }{ m_{\mathrm{prev}} } \right)^{2/3} \cdot k_B T_{\mathrm{prev}} / \varepsilon = \left( \frac{14}{7} \right)^{2/3} \cdot 1.2 = 1.9$$.

However, as the size of the hydrophobic block grows, the transition state should also have an additional contribution due to the free energy cost of passing a globule
through the brush formed by the hydrophilic coronas. So this simple estimate should eventually break down.

Implicit in our arguments is that polymer
transfer accounts for the diffusive behavior in \autoref{fig:polymer_msd}. Therefore, the diffusion
coefficient can be roughly estimated from the time $\tau_{PT}$ it takes a polymer to travel the
nearest neighbor micelle distance $~a_L$. Comparing this estimate with the fit to \autoref{fig:polymer_msd}, a good agreement is obtained: 
\begin{eqnarray}\label{Eq:Diffusive_estimate}
\frac{\langle ({\vec r}(t)-{\vec r}(0))^2 \rangle}{t}&=&1.3\cdot 10^{-5} \frac{\sigma^2}{\Delta t}
\\\nonumber
   \frac{a_L^2}{\tau_{PT}}=a_L^2r_{PT}&=&1.1\cdot 10^{-5}\frac{\sigma^2}{\Delta t} \ .
\end{eqnarray}

\begin{figure}[bt]
\includegraphics[width=8cm]{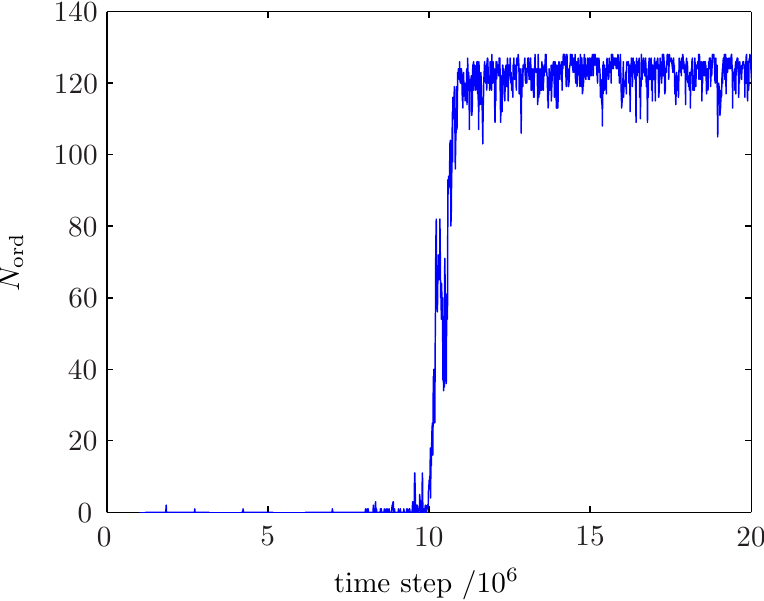}
\caption{\label{fig:n_ordered_micelles} Number of micelles in the ordered phase $N_{\mathrm{ord}}$
as a function of time for a single simulation run of the $A_{10}B_{7}A_{10}$ polymer at
$\phi_P=0.20$ and temperature $k_B T / \varepsilon = 1.2$. The origin of the time axis (t=0) on this plot indicates the time step where the simulation was started from a random configuration.}
\end{figure}

\subsection{Dynamics of micellar crystal formation}

The structure factor $S(\vec{q})$ is sufficient as an order parameter to determine if the entire
system is in an ordered state, but it reveals no information about the dynamics before that final
state is formed. For this, we turn to the bond order analysis~\cite{Steinhardt1983} and apply it
to all the micelles at every time step in simulation runs performed on the $A_{10}B_{7}A_{10}$
polymer with $N_{\mathrm{poly}} = 2134$. In short, at any given time step where a micellar crystal
may have partially formed, the bond order analysis identifies those micelles
belonging to the ordered and the disordered phases.

One simple way to examine the results is to count the number of ordered micelles $N_{\mathrm{ord}}$ in the simulation box at every time step. An example from one simulation run is shown in \autoref{fig:n_ordered_micelles}, plots for all other simulation runs performed are qualitatively very similar, though the ordered phase appears at different times.

After starting the simulation shown in \autoref{fig:n_ordered_micelles} from a random  configuration, micelles quickly form in only a few thousand time steps and single micelle degrees of freedom are equilibrated before one million time steps have passed. The
system then explores configuration space as polymers are transferred and micelles travel anywhere
from a few $\sigma$ up to $100 \sigma$ without any micelles appearing ordered until time step 10
million. At this point, the number of ordered micelles grows very quickly over the next one million
steps until all 128 micelles are in the bcc lattice and remain there for the duration of the
simulation.

During this short time span of $N_{\mathrm{ord}}$ growth, only $7\%$ of the polymers in the box are transferred between
micelles, such a small amount that it cannot fully account for the ordering of all the micelles. During the same time interval, a detailed analysis shows that some micelles move significantly (up to $10\sigma$)
before the ordered phase finishes forming. This may suggest that micellar crystal
formation is a two step process, where first individual micelles are equilibrated by polymer
transfer followed by a second step where polymer transfer becomes irrelevant and the actual crystal
grows via the movement of micelles.

However, some additional simulations performed to test this hypothesis do not support it.  Single micelle degrees
of freedom in these tests are first equilibrated at $k_B T / \varepsilon = 1.2$ for 5 million time steps and then the kinetic temperature
is quenched to $k_B T / \varepsilon = 1.0$ to significantly reduce polymer transfer and allow micelle movement. None of
these simulation runs resulted in the formation of an ordered phase. We therefore conclude that a significant amount of polymer transfer remains a critical component in the actual growth of ordered micellar crystals.

\begin{figure}[hbt]
\includegraphics[width=8cm]{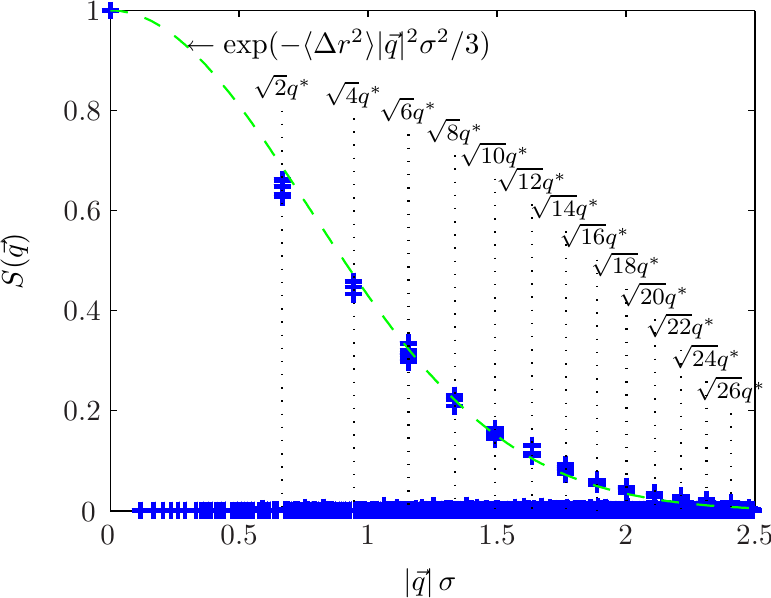}
\caption{\label{fig:bcc_2134_sfactor} Structure factor calculated after the lattice formed for the
$A_{10}B_{7}A_{10}$ polymer simulation run at $\phi_P = 0.20$, $k_B T / \varepsilon = 1.2$, and
$N_{\mathrm{poly}} = 2134$. The full 3D $S(\vec{q})$ is plotted as a scatter plot of
$S(\vec{q})$ versus $|\vec{q}|$. The multiplicity of the various peaks can be seen. Vertical dotted lines indicate
the location of identified peaks, and their positions relative to $q^* = 4\cdot2\pi/L$ are also
noted (the factor of 4 is included because there are 4 unit cells along the box length L).}
\end{figure}

\begin{figure}[hbt]
\includegraphics[width=8cm]{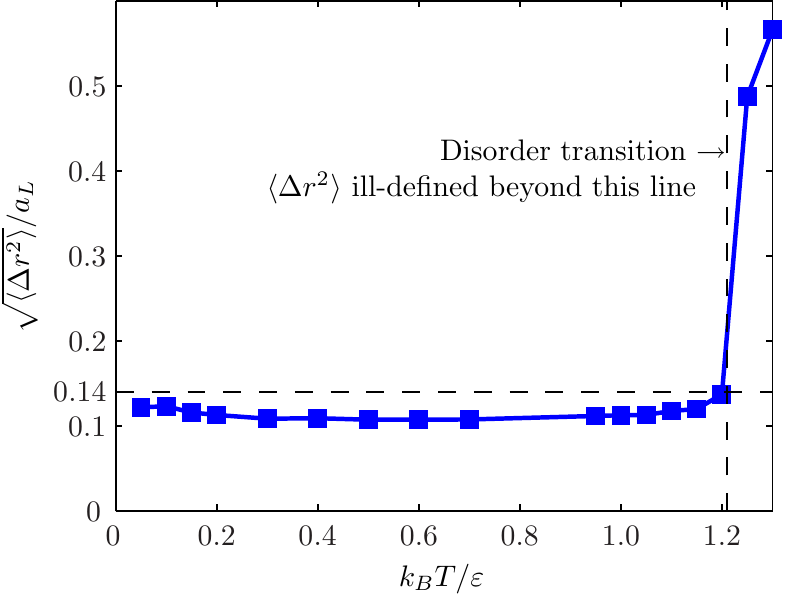}
\caption{\label{fig:lindemann} Fluctuations in micelle positions vs. temperature calculated for a
$A_{10}B_{7}A_{10}$ polymer simulation run at $\phi_P = 0.20$, $k_B T / \varepsilon = 1.2$, and
$N_{\mathrm{poly}} = 900$. The micellar crystal was formed at $k_B T/\varepsilon = 1.2$ and then
cooled to the target temperature without disrupting the lattice. Simulation runs heated to a higher
temperature do disrupt the lattice and $\Delta r^2$ becomes ill-defined. Fluctuations on the y-axis
are plotted as a ratio relative to $a_L$, the nearest neighbor distance in the lattice.}
\end{figure}

\section{Properties of micellar crystals}\label{sec:crystal_static}

\subsection{Structure factor}

We calculate the structure factor (\autoref{Eq_Sq}) for the equilibrium state of every simulation
run performed and use it as an order parameter to determine if a micellar crystal is present. In
those systems where the crystal does form it is a perfect single crystal and, correspondingly, a
large number of peaks can be identified in $S(\vec{q})$ as shown in \autoref{fig:bcc_2134_sfactor}.
Peaks occur in reciprocal space at discrete points $\vec{q} = \vec{G}$, where $\vec{G}$ are the
reciprocal vectors of the corresponding lattice.

Peaks evident in \autoref{fig:bcc_2134_sfactor} decrease in magnitude for larger values of $\vec{G}$. This damping is expected to follow a Debye-Waller factor~\cite{Kittel2005} approximately described as
\begin{equation}\label{Eq:DebyeWaller}
    S(\vec{q}=\vec{G})\propto \exp(-\langle \Delta r^2 \rangle |\vec{G}|^2/3) \ ,
\end{equation}
where $\langle \Delta r^2 \rangle$ is the mean square displacement of the micelle center of mass
from its ideal lattice position. A curve fit, shown in \autoref{fig:bcc_2134_sfactor}, is in
excellent agreement with \autoref{Eq:DebyeWaller}. The Lindemann ratio $f_L$ is defined as
\begin{equation}\label{Eq:Lindemann}
    \sqrt{\langle \Delta r^2 \rangle}=f_L a_L
\end{equation}
where $a_L$ is the nearest neighbor distance between micelles. Using the parameters of the curve fit yields $f_L\approx 0.14$.

The Lindemann parameter $f_L$ can alternatively be computed directly by measuring the mean square displacement of micelles about their average lattice positions. The results, shown in \autoref{fig:lindemann} agree remarkably well with the estimate from the Debye-Waller factor. Furthermore, it has been established empirically that approximately at $f_L = 0.13$ solids melt into disordered states~\cite{Hansen1976}, a result that is also supported from our simulations as we observed no micellar crystals form at kinetic temperatures greater than $k_B T / \varepsilon = 1.2$

\begin{figure}[hbt]
\includegraphics[width=8cm]{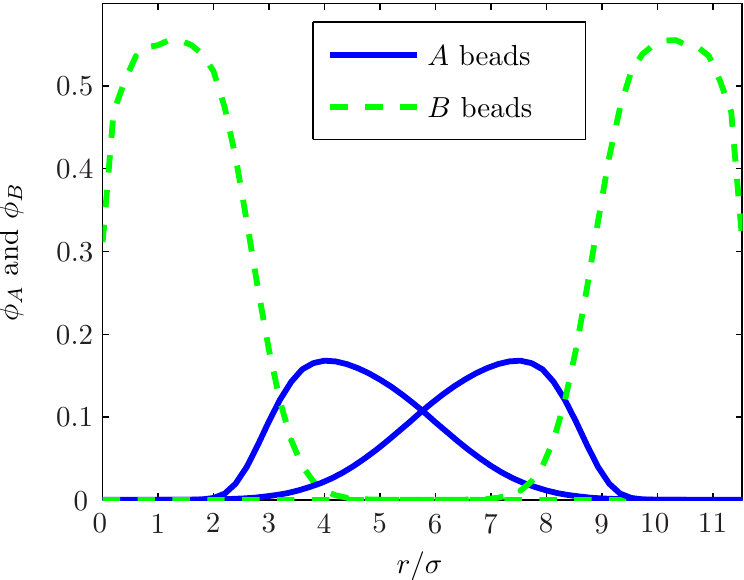}
\caption{\label{fig:mic_overlap} Radial density distribution for two nearest neighbor micelles
superimposed with a separation of the nearest neighbor distance in the bcc lattice. The y-axis
plots the volume fraction of the different beads belonging to a micelle in the local environment
around the micelle. The results were calculated from a $A_{10}B_{7}A_{10}$ simulation run at
$\phi_P = 0.20$, $k_B T / \varepsilon = 1.2$, $N_{\mathrm{poly}} = 2134$ and averaged over all
micelles and time steps after the micellar crystal has formed.}
\end{figure}

\subsection{Structure of the lattice of micelles}

Micelles in the $A_{10}B_{7}A_{10}$ polymer system are arranged with the centers of mass sitting on
a bcc lattice with a nearest neighbor spacing of $a_L = 11.53\sigma$. This number remarkably agrees
with the length of the polymer if stretched completely into a straight line across the diameter of
a circle from the center of one nearest neighbor micelle to the opposite one, $N_{\mathrm{mon}}
\cdot r_0 = 27 \cdot 0.83\sigma = 22.41\sigma \sim 2 a_L$, where $r_0 = 0.83\sigma$ is the
equilibrium bond length. This suggests that polymers are maximally stretched, either along the
diameter or in a bent configuration. Detailed visual examinations of simulation
snapshots indicate that there are a significant number of polymers in the bent configuration, but there are also some in the linear extended
configuration. The micellar cores are liquid-like, and over time, a given polymer constantly
switches from linear to bent configurations.

The previous arguments also suggest a significant amount of overlap of the coronas between two
neighboring micelles in the lattice. To examine this more quantitatively we calculate the
micelle density distribution as a function of the radius averaged over all micelles and all time
steps in the simulation after the micellar crystal has formed. \autoref{fig:mic_overlap} shows
the results, confirming the overlap. Hydrophilic $A$ beads from one micelle explore the solvent
until occasionally bumping into the hydrophobic core of one of the nearest neighbor micelles.

A remarkable aspect of micellar crystals found in this work is the stability of the average aggregation number. If the
polymers within the micelle are maximally stretched, the core of the micellar radius is
$R_c=mr_0\sigma/2$, so the aggregation number can be estimated from
\begin{equation}\label{Eq:Navv_StrongStretched}
  \langle N_{agg} \rangle = \frac{4\pi(m r_0/2)^3\sigma^3}{m r_0 \sigma^3}=\frac{\pi}{6} m^2 r^2_0
  \ .
\end{equation}
For the $A_{10}B_{7}A_{10}$ this yields $ \langle N_{agg} \rangle=17.7$, in good agreement with
the simulation results in \autoref{tab:N_test}. Aggregation numbers for the other polymers simulated 
do not agree, implying that the polymers in those systems are not maximally stretched. 

\section{Conclusions}\label{sec:conclusions}

\subsection{Summary of results}

Cubic micellar crystals of $A_nB_mA_n$ polymers form in MD simulations at sufficiently high concentrations. In order to form the crystals, high enough kinetic temperatures are needed to enable polymer transfer between micelles, which is critical for equilibrating the system.  The polymer transfer process is activated and described by transition state theory. It results in an apparent diffusive behavior of the individual polymers while the lattice of micelles remains stable. Excessive polymer transfer at even higher kinetic temperatures triggers a disorder phase transition to a micelle liquid.

\begin{figure}[tb]
\includegraphics[width=8cm]{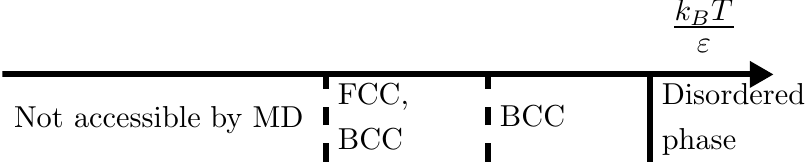}
\caption{\label{fig:mic_phase_diag} Summary of the phase diagram encompassing all simulated
triblocks $A_nB_mA_n$ (all forming cubic phases with long range order).  Near the disorder
transition, bcc is always favored and fcc lattices only begin to appear at lower temperatures. At
even lower temperatures polymer transfer becomes negligible and MD would require prohibitively long
simulations to reach equilibrium. }
\end{figure}

In the process of forming the ordered phase, the system spends a long time in a micellar liquid phase equilibration period where no ordered nucleates are present. Eventually, a large nucleation event takes place and the micellar crystal then grows very quickly, filling the entire simulation box in a relatively short time span. During this growth period, polymer transfer and movement of micelles are both crucial in the formation of the final micellar crystal. The preferred ordering near the disordered transition for all triblocks studied in this system is the bcc lattice. Only at lower kinetic temperatures and for polymers with short hydrophilic groups ($A_6B_7A_6$ at $k_B
T/\varepsilon=1.07$) is there some evidence for a stable fcc phase. These results are summarized in
\autoref{fig:mic_phase_diag}.

All micellar crystals obtained in this work are perfect single crystals displaying a high degree of order. Even in a periodic simulation box with four unit cells along a side, a single additional micelle can disrupt the resulting lattice significantly. The number of micelles is controlled by changing the number of polymers in the box, as the system displays a remarkable stability in the average aggregation number of the micelles that form. In the equilibrium lattice, micelles are closely packed with a significant amount of overlap between the coronas of neighboring micelles. Polymers in the micelles are highly stretched across the liquid hydrophobic core through the solvent  and qualitatively well described within the strongly stretched approximation \cite{Semenov1985,Grason2007}.

\subsection{Stability of the bcc lattice near the disorder transition}

Our simulations results show a strong preference for bcc lattices near the disorder transition. A
similar result has been experimentally observed for PS-PI diblocks \cite{Lodge2002}, where it has
been attributed to the fact that near the disordered phase, micelle aggregation numbers are small.
The phase diagram of f-star polymer systems shows that fcc lattices are only stable for large
number of arms $f>60$, while bcc lattices are favored when the number of arms is small~\cite{Watzlawek1999}. By
considering polymeric micelles as f-star polymers, where the number of arms $f$ is given by
$f\sim 2 \langle N_{agg} \rangle$ then bcc lattices are favored when the aggregation numbers
are small $\langle N_{agg} \rangle \lesssim 30$. This argument, however, hinges on two key
assumptions: the dynamic nature of micelles does not play a significant role and that the
hydrophilic blocks are sufficiently long. The first assumption is already somewhat problematic given the importance of polymer transfer found in this work. As
for the second assumption, a criteria establishing how long hydrophilic blocks should be has been
put forward in Ref.~\cite{McConnell1996}, where it is shown that if the size of the corona is $L_c$
and the core radius $R_c$, bcc lattices are favored for $L_c/R_c > 1.5$. Our results for the
$A_{10}B_7A_{10}$ and $A_{6}B_7A_{6}$ yield $L_c/R_c \sim 0.9$ and $L_c/R_c \sim 0.7$.  It is
therefore not possible to attribute the stability of the bcc lattices observed in our simulations
as being a consequence of the small aggregation numbers $\langle N_{agg} \rangle \lesssim 30$.

It is tantalizing to interpret the stability of the bcc lattice in terms of the Alexander-McTague
(AM) scenario \cite{Alexander1978}, where it was argued that bcc should be generally expected to be
the stable phase near a (weakly first order) disorder transition. Subsequent analysis however,
showed that cubic lattices other than bcc cannot be ruled out near the disorder
transition~\cite{Groh1999,Klein2001}. In Ref.~\cite{Groh1999,Klein2001} it is shown that the
characteristic property of bcc lattices is that their free energy is closer to the disordered state,
thus suggesting that bcc lattices follow an Ostwald step rule \cite{Ostwald1897}, where the solid
phase that nucleates first is the one whose free energy is closest to the disordered (or fluid)
state. In this case, the complete crystallization process would require an additional step, where
after the bcc crystallites are formed, they gradually evolve towards the stable thermodynamic
phase. Certainly, it follows from our results that fcc lattices are difficult to obtain by MD for the systems we simulate, but
at least in one system ($A_6B_7A_6$ at $k_B T/\varepsilon=1.07$) the fcc lattice has been obtained
reproducibly, and did not proceed through an intermediate bcc step. In addition, for this very same
system, closer to the disordered state ($k_B T/\varepsilon=1.1$), no fcc structure was found to be
stable. Although not completely conclusive, our results are more consistent with the bcc as being a
stable thermodynamic phase near the disordered phase.

There are serious limitations in identifying micelles as simple particles,
because as the disordered phase is approached micelle aggregation numbers decrease and polymers
become essentially free. So the disordered phase is not a simple liquid as it is assumed by AM and
all subsequent work. Similar analysis in polymer melts \cite{Leibler1980}, shows that in the
vicinity of the spinodal, the only possible phase with cubic symmetry is bcc. Beyond the
spinodal, other structures are favored \cite{Marques1990}. Based on the previous discussion, we
attribute the stability of the bcc phases observed in our simulations as a reflection of the
admittedly non-rigorous statement that bcc phases are \emph{usually favored} near the disordered
transition. We defer to future work to establish this result within a rigorous framework, where all
the nuances involved in micelle formation are properly taken into account.

\subsection{Implications for Pluronic systems}

The $A_{10}B_{7}A_{10}$ and $A_{20}B_{14}A_{20}$ polymers discussed in this paper provide
coarse-grained descriptions of Pluronic F127. All simulations have been carried out in very good
solvent conditions for the $A$ beads and at total volume fractions of 15--20\%. Experimental
results in this region of the phase diagram are surprisingly disparate. Simple
cubic~\cite{Prud1996}, bcc~\cite{Mortensen1995} and fcc~\cite{Liu2000} have all been proposed as
the structure in this region. Very recently, on the basis of new experimental data, the situation
has been thoroughly reviewed by Li et al.~\cite{Li2006} (although for significantly higher
temperatures), see also Ref.~\cite{Pozzo2007}, but without clear conclusive results. Our
theoretical analysis clearly favors the bcc lattice close to the disorder transition. The
comparison of our results with the experimental F127 system is more accurate at low temperatures
(at 20-25 C), where the water can be considered as a good solvent for PEO, as discussed in
Ref.~\cite{Anderson2006}.

\subsection{Outlook}

We have shown that MD allows a detailed investigation of both the dynamics as well as the
thermodynamic equilibrium of micellar crystals. Many studies have been performed by modeling
micelles as point particles, where the complex structure of the micelles is accommodated through
refined two-body potentials, either derived analytically or empirically. While successful in many
situations, two-body potentials do not account for the dynamic nature of micelles, which play a
critical role in determining the phases of the system, particularly near the disorder transition.

There are a few areas where further work needs to be performed. First, all simulations in this work
have been performed in good solvent conditions. MD with implicit solvents of different quality are
also of great interest, especially to determine the phases of Pluronic systems over a wide range of
temperatures ~\cite{Anderson2006}. Next, the largest micellar crystal formed in this work contains
2134 polymers (57,618 beads). We were unable to find any order in larger systems, even after
running as many as 50 million time steps. It is possible that significantly longer simulations may
be required for larger systems. Also, the range of applicability of MD is restricted to a
relatively narrow range near the disorder transition, as schematically shown in
\autoref{fig:mic_phase_diag}. Finally, the role of the solvent will need to be investigated in more
detail, as it is found in the Rouse vs.~Zimm dynamics for simple homopolymers~\cite{Doi1986}. Future
studies will be necessary to completely clarify and expand all these issues.

The relevance of our study goes beyond pure systems. In Ref.~\cite{Enlow2007} for example, Pluronic
polymers have been used to template the growth of an inorganic phase of calcium phosphate, aimed at
creating new polymer nanocomposites with lightweight/high strength properties or that mimic the
structure of real bone. An understanding of the pure systems is clearly a prerequisite for accurate
models of polymer nanocomposites. We hope to report more on this topic in the near future.

\begin{acknowledgments}
We acknowledge interest and discussions with S. Mallapragada, K. Schmidt-Rohr and G. Grason. We also
thank J. Schmalian for clarifications regarding the Alexander-McTague paper. This work is supported
by DOE-BES through the Ames lab under contract no. DE-AC02-07CH11358 and partially supported by the
NSF through grant DMR-0426597. \end{acknowledgments}

\end{document}